\documentclass[aps,pre,reprint,superscriptaddress]{revtex4-1}

\usepackage{url}
\usepackage{color}
\usepackage{graphicx}
\usepackage{bm}
\usepackage{amsmath}
\usepackage{mathtools}
\usepackage{amssymb}
\usepackage{cleveref}
\usepackage{physics}
\usepackage{verbatim}

\begin{document}


\title{Twisted loxodromes in spindle-shaped polymer nematics}


\author{Helen S. Ansell}
\email{ansellh@sas.upenn.edu}
\author{Randall D. Kamien}
\email{kamien@physics.upenn.edu}
\affiliation{Department of Physics and Astronomy, University of Pennsylvania, Philadelphia, Pennsylvania 19104, USA}



\date{May 24, 2021}

\begin{abstract}
We develop an energetic model that captures the twisting behavior of spindle-shaped polymer microparticles with nematic ordering, which display remarkably different twisting behavior to ordinary nematics confined to spindles. We have previously developed a geometric model of the twisting, based on experimental observations, in which we showed that the twist pattern follows loxodrome spirals  [Ansell \textit{et. al., Phys. Rev. Lett.}, \textbf{123}, 157801 (2019)]. In this study, we first consider a spindle-shaped surface and show that the loxodrome twisting behavior of our system can be captured by the Frank free energy of the nematic with an additional term constraining the length of the integral curves of the system. We then extend the ideas of this model to the bulk and explore the parameter space for which the twisted loxodrome solution is energetically favorable.
\end{abstract}


\maketitle


\section{Introduction}
	Spontaneous twisting is observed in many liquid crystalline systems in which the constituents are achiral and the confinement does not require a twisted configuration~\cite{Press1974, Pang1994, Volovik1983, Lavrentovich1990, Drzaic1999, Jeong2014,  Jeong2015, Yang2008, Nayani2017, Tortora2011, Vanzo2012}. Such behavior may be observed in systems for which the relative magnitudes of the elastic constants of the material make the twist deformations energetically favorable~\cite{ Yang2008, Jeong2014, Jeong2015, Lavrentovich1990, Drzaic1999,Nayani2017, Tortora2011, Williams1985-028, Williams1986, Prinsen2004}. Examples of such spontaneous twisting have been observed in achiral nematic molecules confined to cylindrical capillaries~\cite{Jeong2015} and droplets with homeotropic anchoring~\cite{Yang2008}. Twisted bipolar structures in spherical nematic droplets provide a frequently observed example of this phenomenon~\cite{Volovik1983, Lavrentovich1990, Drzaic1999, Jeong2014}. Theoretical work by Williams~\cite{Williams1986} showed that the twisted bipolar configuration is energetically favorable in spherical bipolar droplets if the elastic constants of the material satisfy the inequality \(K_2<K_1 - 0.43 K_3\), where \(K_1\), \(K_2\) and \(K_3\) are respectively the splay, twist and bend elastic constants of the nematic. More recently, twisted bipolar structures have also been observed in elongated spindle-shaped nematic droplets~\cite{Tortora2011, Vanzo2012, Nayani2017} and polymer liquid crystalline microparticles~\cite{Wang2016, Ansell2019}. 
		
		Understanding the behavior of nematics confined to elongated spindle-shaped regions has long created interest due to the spindle-shaped tactoids that form in lyotropic liquid crystals as the nematic phase nucleates. These tactoids were first observed by Zocher in the 1920s~\cite{Zocher1925} in vanadium pentoxide and have since been observed in a host of inorganic and biological materials~\cite{Bernal1937, Bernal1941, Zocher1960, Puech2010, Kim2013, Modlinska2015, Jamali2015, WangPX2016, Nayani2017, Nystrom2018}. The tactoids consist of regions of nematic that coexist with the surrounding isotropic phase. The prevalence of tactoids in lyotropic systems, as well as bipolar structures in thermotropic droplets, has inspired many studies aiming to understand the shape and director field of such systems~\cite{ Kaznacheev2002, Kaznacheev2003, Kalugin1998, Williams1985-028, Williams1986, Bates2003, Prinsen2003, Prinsen2004b, Prinsen2004, Tortora2011, Vanzo2012, vanBijnen2012, Metselaar2017, Safdari2021}, which depends on the interplay between the elasticity, surface tension and droplet size as well as the effect of applied fields. Using scaling arguments, Prinsen and van der Schoot~\cite{Prinsen2003, Prinsen2004b} showed that the director field configuration in a tactoid depends on its size, with smaller tactoids having a homogeneous director field while larger tactoids have a quasi-bipolar director field that becomes exactly bipolar in the infinite volume limit. The same authors also showed that spindle-shaped bipolar tactoids with pointed tips have a lower free energy than comparable prolate spheroidal-shaped tactoids~\cite{Prinsen2003}, consistent with the spindle shapes observed experimentally~\cite{Zocher1925, Bernal1941, Zocher1960, Kaznacheev2002, Kaznacheev2003, Tortora2011, Puech2010, Kim2013,Nystrom2018, Modlinska2015, Jamali2015}.
			
	Inspired by the work of Williams~\cite{Williams1986}, Prinsen and van der Schoot also investigated twisted bipolar structures in spindle-shaped tactoids~\cite{Prinsen2004} and generalized the Williams inequality to account for the anisotropic shape of the spindle. They showed that the maximum value of the twist elastic constant, relative to the splay and bend constants, at which twisting is preferable decreases as the tactoids become smaller in volume, and consequently more elongated. The typical values of elastic constants in many lyotropic systems do not satisfy the requirements for twisting, so these systems would not be expected to exhibit a twisted bipolar configuration~\cite{Prinsen2004}. However, twisted bipolar structures have been observed in lyotropic chromonic liquid crystals~\cite{Tortora2011, Jeong2014, Nayani2017}, a class of materials in which the twist elastic constant is significantly smaller than the splay and bend elastic constants. These materials therefore appear to satisfy the inequality required for twisting and are indeed observed to display highly twisted bipolar configurations. 
			
Recently, we investigated the twisting behavior of spindle-shaped polymer liquid crystalline microparticles~\cite{Ansell2019} and developed a geometric model to describe their twisting behavior. These bipolar polymer particles were created by polymerizing spherical bipolar nematic droplets containing the reactive mesogen RM257 at low wt\% in a mixture with the non-reactive liquid crystal 5CB. After removing the 5CB, the initially spherical polymer particles deswell anisotropically in solvents into elongated spindle shapes, forming a chiral twisted bipolar structure in the process. In our geometric model, we showed that the twisting behavior of the polymers on the surface was well-described by a type of spiral called a \textit{loxodrome}, in which the angle between the integral curves of the system and the principal directions of the surface are the same at every point along the curve. Such a twisting structure has been previously assumed in theoretical studies of twisted bipolar structures~\cite{Williams1986, Prinsen2004} and it has been suggested that such structures are consistent with observations of other twisted bipolar systems~\cite{Volovik1983, Tortora2011, Vanzo2012}. 

	While our polymer system displays a twisted bipolar structure that is well-described by loxodromes, the overall twisting behavior is not consistent with that predicted for spindle-shaped tactoids~\cite{Prinsen2004}. Our system is consistent with the model in that we observe that smaller volume spindles have a larger aspect ratio. However, we also observe that larger aspect ratio spindles are \textit{more} twisted than those that are closer to spherical, in direct contrast with the tactoid model. In this study, we therefore develop an energetic model that captures the behavior of these twisted spindle-shaped polymer particles. Given the nematic ordering of the polymer system, we base our model around the Frank free energy of the nematic and seek to incorporate additional terms to capture the polymer behavior. We show that, as was the case in our geometric model, incorporating a constraint on the length of the integral curves in our system results in twisted loxodrome solutions that minimize the free energy and predict behavior that is consistent with our previous experimental observations.	
	
	The structure of the remainder of this paper is as follows. In section 2, we introduce the parameterization of the spindle shapes and discuss the Frank free energy that will be the starting point of our model. In section 3 we consider a nematic confined to a spindle-shaped surface. We show that twisted loxodrome solutions do not minimize the Frank free energy, but that adding an additional length constraint term allows for an exact twisted loxodrome solution if \(K_1 = K_3\). We show that the loxodrome solution is a good approximation in cases where the deviation between \(K_1\) and \(K_3\) is small. In section 4 we consider the conditions on the shape profile of a surface of revolution for it to support a twisted loxodrome solution, and show that the most general surface that supports such a solution is the general torus. In section 5 we then turn our attention to the bulk and show that, while not exact, twisted bispherical loxodrome solutions are a good approximation to the bulk solution. We use this to extend the ideas of our geometric model to the bulk structure and show that the resulting model favors twisting for a wider parameter range than expected in twisted nematic systems that do not have the additional length constraint.

\section{System parameterization and energy}					
In this study, we consider a nematic system confined within a spindle-shaped droplet with strong planar anchoring at the surface in which the nematic director field adopts a bipolar configuration. We define a spindle as a surface of revolution formed by revolving a minor arc of a circle about the chord connecting its endpoints. In line with previous investigations, we take the director field within the bipolar droplets to follow the symmetries of a bispherical director field~\cite{Williams1985-028, Williams1986, Kaznacheev2002, Kaznacheev2003, Prinsen2003, Prinsen2004}, which was shown by Williams~\cite{Williams1985-028} to be a good approximation for the numerically solved director field in the one constant approximation. The internal bispherical structure can be considered to consist of layered spindles of the same major axis length and with different minor axis lengths, arranged so that the tips of all of the spindle layers coincide at the locations of the two boojum defects.

\begin{figure}
		\centering
		\includegraphics[width=0.45\textwidth]{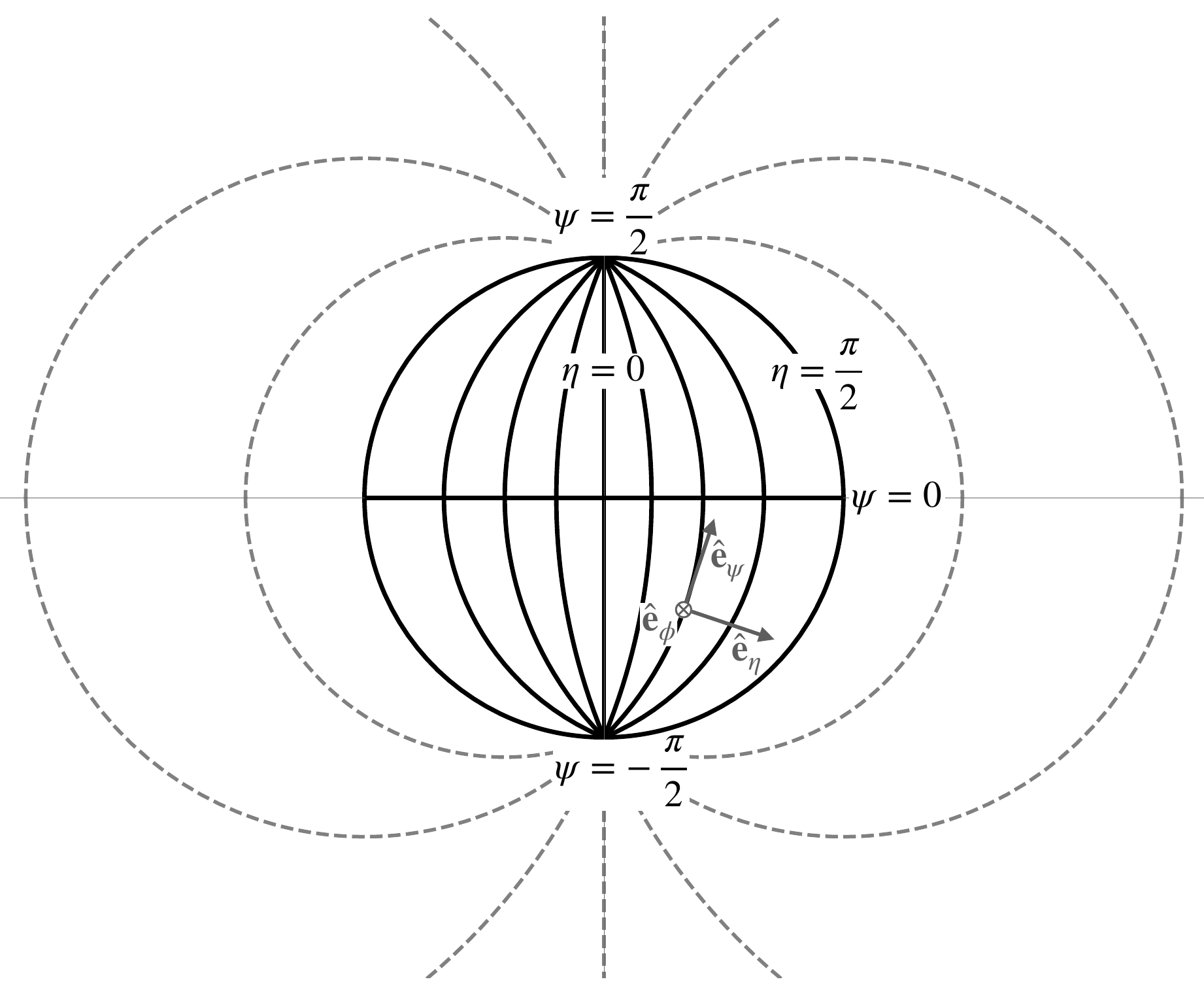}
		\caption{Schematic slice through the bispherical coordinates system. Black arcs and dashed gray lines indicate surfaces of constant \(\eta\), which all pass through the two poles of the structure. A complete surface of constant \(\eta\) is formed by revolving one of these arcs about the line connecting the poles. Positions on each spindle surface (fixed \(\eta\)) are described in terms of the angle \(\psi\), which varies between \(0\) at the equator and \(\pm \pi/2\) at the tips, and \(\phi\) which is the azimuthal angle.}
	\label{fig:bisph-schem}
	\end{figure}

The structure of the spindles naturally leads us to describe positions in the system using bispherical coordinates \((\eta, \phi, \psi)\), with corresponding orthonormal unit vectors \((\vu{e}_{\eta}, \vu{e}_{\phi}, \vu{e}_{\psi})\), which are shown in \cref{fig:bisph-schem}. These coordinates can be related to Cartesian coordinates with a common origin through the transformation
\begin{equation}
	\mqty(x \\ y \\ z) = \frac{1}{Z}\mqty(\sin{\eta}\cos{\psi}\cos{\phi} \\ \sin{\eta}\cos{\psi}\sin{\phi} \\ \sin{\psi}),
\end{equation}
where \(Z = 1+\cos{\eta}\cos{\psi}\). In the bispherical coordinate system, a surface of constant \(\eta\) defines a spindle-shaped surface. The value of \(\eta\) varies within a spindle between zero along the line connecting the two boojum defects and a maximum value \(\eta_0\) at the spindle surface. The aspect ratio \(u_0\) of the spindle is \(u_0 = (1+\cos{\eta_0})/\sin{\eta_0} =  \cot{(\eta_0/2)}\), which has a minimum value of one for a sphere, for which \(\eta_0 = \pi/2\), and increases as \(\eta_0\) decreases and the spindle becomes more elongated. The coordinate \(0\leq \phi < 2\pi\)  is the azimuthal angle while \(\psi\) represents the polar angle. We choose the origin of \(\psi\) such that it varies between zero at the equator of the spindle and \(\pm \pi/2\) at the tips. 

The free energy of the nematic can be described using the Frank elastic free energy~\cite{Frank1958} 
			\begin{align}
				&F_F = \frac{1}{2}{\int\dd V}\left[K_1 (\vb{n}\, \div\vb{n})^2 + K_2(\vb{n}\vdot\curl{\vb{n}})^2 + \right. \nonumber\\
				&\left.\!\!\! K_3(\vb{n}\cp\curl{\vb{n}})^2 	\right]				
					- K_{24} {\int \dd \vb{S}}{\vdot}\left[ \vb{n}(\grad{\vdot}\vb{n}) - (\vb{n}{ \vdot }\grad)\vb{n}\right]
			\label{eq:Frank}
 			\end{align}		
where \(\vb{n}\) is the spatially varying nematic director field that has the property \(\vb{n} = - \vb{n}\). The elastic constants \(K_1\), \(K_2\) and \(K_3\) are, respectively, associated with the splay, twist and bend deformations of the director field within the volume \(V\) of the droplet. The \(K_{24}\) term is integrated over the surface \(S\) of the droplet and describes saddle-splay deformations.

The elastic constants must satisfy the Ericksen inequalities~\cite{Ericksen1966}, which require that \(K_1\), \(K_2\) and \(K_3\) are non-negative while \(K_{24}\) must satisfy \(\abs{K_{24}}\leq 2 \text{ min}(K_1,K_3)\). The sign of \(K_{24}\) changes the director field curvature preferred by the saddle splay term. Recall that when the director field is tangent to a surface, positive \(K_{24}\) leads to a configuration in which positive Gaussian curvature is preferred while negative \(K_{24}\) prefers a saddle configuration. The saddle configuration is not axially symmetric and is therefore incompatible with our expectation that our twisted solutions display axial symmetry. We therefore expect that our system must have a positive \(K_{24}\) solution. 
	
Due to the mathematical complexity of the equations required to describe the behavior of twisted bipolar nematic spindles, simplifications must be made in order to make progress. In the bulk, the twist angle is expected to increase from zero along the central axis of the spindle to a maximum value on the surface~\cite{Williams1986, Prinsen2004}. In the bispherical structure the director field is tangent to surfaces of constant \(\eta\). Further assuming that the twist angle, measured as the angle between the director field and the surface meridian, is constant on a surface of constant \(\eta\) therefore gives a director field that depends only on the \(\eta\) coordinate. Using these assumptions, the resulting nematic director field has a layered spindle structure in which the director field on each surface of constant \(\eta\) follows loxodrome spirals at a twist angle determined by the value of \(\eta\). Williams~\cite{Williams1986} used these assumptions when considering the effect of the values of the bulk elastic constants on the transition to a twisted state in a bipolar spherical droplet. Following on from this, Prinsen and van der Schoot~\cite{Prinsen2004} used these same assumptions to consider the conditions for twisting in elongated spindle-shaped structures as well as director fields for which the boojum defects are virtual and sit outside of the droplet surface. The assumption was justified as being due to the expectation that the twist angle would vary more with \(\eta\) than it would vary with \(\psi\) on a surface of constant \(\eta\)~\cite{Prinsen2004}. 

While describing a director field in terms of loxodromes is mathematically convenient, due to the lack of \(\psi\) dependence, it is not necessarily intuitive that the system would choose to adopt this twisting structure. Indeed, close to the spindle tips such a structure creates a region of high twist. In general, loxodromes are not geodesics of a surface and do not minimize the curvature of the director field. From Clairaut's relation~\cite{doCarmo}, we can determine that on a spindle surface the only loxodromes that are also geodesics are those for which the director is parallel to the meridians of the surface, which gives an untwisted bipolar configuration. We therefore first approach the question of when a twisted loxodrome solution is favorable in our spindles by considering the conditions under which the loxodrome twisting pattern is an exact energy minimum on a spindle-shaped surface.

\section{Loxodromes solutions on spindle surfaces}
We consider a thin spindle-shaped shell of nematic with the director field tangent to the surface everywhere and investigate the conditions under which a twisted loxodrome structure minimizes the free energy. We assume the director field follows the azimuthal symmetry of the surface and is therefore independent of \(\phi\). As such, 
	\begin{equation}
		\vb{n} = \cos{\beta(\psi)}\vu{e}_{\psi} +  \sin{\beta(\psi)}\vu{e}_{\phi} 
		\label{eq:n-surf}
	\end{equation}	
	describes a general director field on the surface. Here \(\beta(\psi)\) is the angle between the director field and the \(\vu{e}_{\psi}\) direction on the surface. The nematic symmetry of the system means that \(\beta(\psi)\) and \(\beta(\psi)+\pi\) are equivalent while the symmetry of the spindle shape means that \(\beta(\psi)\) and \(2\pi-\beta(\psi)\) are energetically equivalent states so that left- and right-handed twisting are equally likely. We can therefore take \(\beta(\psi)\) to be in the range \(0\leq\beta(\psi)\leq\pi/2\) without loss of generality.
	
	We wish to determine whether or not a twisted loxodrome structure minimizes the Frank free energy given in \cref{eq:Frank}. We note that on the spindle surface the twist term is zero. The loxodrome solution obeys \(\beta(\psi) = \beta_0\), where \(\beta_0\) is a constant, and \(\beta'(\psi) = \beta''(\psi) = 0\). In the one-constant approximation, in which \(K_1 = K_3\), in the limit that the spindle is a perfect sphere (\(\eta_0 = \pi/2\)), the saddle splay term is zero and loxodrome solutions of arbitrary twist angle minimize the free energy. However, allowing either \(K_1\neq K_3\) or \(\eta_0 \neq \pi/2 \) removes this solution. In this case, the only constant angle solutions are \(\beta_0 = 0, \pi/2\), which correspond to untwisted configurations with the director parallel to \(\vu{e}_{\psi}\) and \(\vu{e}_{\phi}\) respectively. We therefore find that the Frank free energy alone does not support twisted loxodrome solutions on spindle-shaped surfaces. 
	
	In our geometric model of twisted spindle-shaped polymer particles with nematic ordering, we showed that the twisting pattern was well-described by loxodromes~\cite{Ansell2019}. A key feature of our model was the introduction of a constraint on the length of the integral curves of the twist pattern. We therefore introduce a length constraint term into our free energy and determine whether this addition allows a twisted loxodrome solution. A constraint on the length of the integral curves of the director field takes the form 
	\begin{equation}	
		F_l = \gamma \left(\int \dd s - l_0\right)
	\end{equation}
	where \(\gamma\) is a Lagrange multiplier, \(\dd s\) is the line element and \(l_0\) is the fixed curve length.	A loxodrome of twist angle \(\beta_0\) on the spindle surface connecting the two tips has length \(l_{\beta_0} = 2 \eta_0\csc{\eta_0}\sec{\beta_0}\). Taking \(\gamma\) to be positive, this term therefore works to minimize the curve length and therefore the twist angle on the surface.
	
	We seek twisted loxodrome structures that minimize our new total free energy \(F = F_F + F_l \), where the general form of the director field is again given by \cref{eq:n-surf}. The resulting Euler-Lagrange equation that \(\beta(\psi)\) must satisfy is 
	\begin{align}
			0&=  K_m \sin{\eta_0}[\beta'(\psi)\sin{\psi} - \beta''(\psi)\cos{\psi} ] \nonumber \\ 
			&- K_{24}\cos{\eta_0} \frac{\sin{2 \beta(\psi)}}{Z}	+\frac{\gamma}{2\pi}\frac{\sec{\beta(\psi)}\tan{\beta(\psi)}}{Z} \nonumber \\
			& -  \Delta K\sin{\eta_0}\cos{\psi}\biggl[\beta'(\psi )^2 \sin{2\beta(\psi)} +
						[\tan{\psi}\beta'(\psi) \nonumber \\ &
						-\beta''(\psi)]\cos{2\beta(\psi)}+\frac{\sec^2{\psi}\sin{2\beta(\psi)}}{Z}\biggr],
			\label{eq:eeq-beta(psi)}	
	\end{align}		
	where we have introduced \(K_m = (K_1+K_3)/2\) and \(\Delta K = (K_1-K_3)/2\). After substituting the conditions for a loxodrome solution into this expression, the final \(\Delta K\) term still has \(\psi\) dependence. We therefore find that a twisted loxodrome solution is possible only if \(\Delta K = 0\), which corresponds to the one-constant approximation. In this case, the loxodrome twist angle satisfies
		\begin{equation}
			|\cos^3{\beta_0}| = \frac{\gamma}{4\pi K_{24}\cos{\eta_0}}
			\label{eq:cos3beta}
		\end{equation}
		on a spindle with \(\eta_0<\pi/2\). From this expression, we can deduce that \(\gamma\) and \(K_{24}\) must have the same sign. Given that \(\gamma\) is positive by construction, we find that \(K_{24}\) must also be positive. This is consistent with our previous discussion on the expected behavior of \(K_{24}\) for the spindle. 
		
	Whether or not the loxodrome solution is realizable depends on the interplay between \(\gamma\), \(K_{24}\) and \(\eta_0\). If \(\gamma/K_{24}>4\pi\) then a twisted loxodrome solution cannot occur. However, if \(\gamma/K_{24} < 4\pi\) the onset of twisting is determined by a critical value of \(\eta_0\) below which this solution is physically realizable. In this framework we would therefore expect that for a system with fixed \(\gamma/K_{24}<4\pi\), that there is a critical spindle aspect ratio at which the onset of twisting occurs. Spindles with aspect ratios smaller than this critical value (larger \(\eta_0\)) will display an untwisted state while those with larger aspect ratios (smaller \(\eta_0\)) can exhibit twisting. The twist angle on the surface increases with the spindle aspect ratio, which is consistent with our previous observations in twisted nematic polymer particles~\cite{Ansell2019}.
	
	This twisted loxodrome solution is energetically favorable over the untwisted configuration if \(\Delta F = F(\beta_0) - F(0)\), the difference in free energy between the twisted loxodrome solution and the untwisted solution, is negative. We find that \(\Delta F < 0\) whenever the system parameters allow the twisted solution to be possible, meaning that the loxodrome solution is always energetically favorable over the untwisted solution. By numerically solving \cref{eq:eeq-beta(psi)} using the Matlab ODE15s solver~\cite{Shampine1997} with \(\Delta K = 0\) for a range of choices of \(\gamma/K_{24}\) and \(\eta_0\) with boundary conditions setting \(\beta'(0) = 0\) and \(\beta(0)\) to a chosen value, we have verified that the twisted loxodrome solution \(\beta(0) = \beta_0\) is the global minimum of the free energy if the choice of parameters allows that solution to exist.

We now allow the value of \(\Delta K\) to be nonzero and consider the effect this has on the solutions of \cref{eq:eeq-beta(psi)}. Examining the \(\Delta K\) dependent terms in \cref{eq:eeq-beta(psi)}, we observe that final term depends on \(\sec{\psi}\). While well-behaved at the spindle equator, near to the spindle tips this term diverges and dominates the entire expression. Our problem therefore becomes a boundary layer problem in which we must separately consider solutions near to the equator and tips of the spindle.

We first consider the behavior near to the equator, where the \(\Delta K\)-dependent terms do not diverge. We capture the key behavior of the system by performing a regular perturbation expansion in which we seek a solution of the form \(\beta^{\epsilon}(\psi) = \beta^0(\psi) + \epsilon \beta^1(\psi)\), where \(\epsilon\) is a small parameter. We substitute this solution into \cref{eq:eeq-beta(psi)} and expand in powers of \(\epsilon\). Gathering the leading order terms, those for which there is no \(\epsilon\) or \(\Delta  K\) dependence, we find that the twisted loxodrome solution in \cref{eq:cos3beta} is a leading order solution, so \(\beta^0(\psi)\ = \beta_0\). In order to calculate the first correction term, we identify the small parameter \(\epsilon\) with \(\Delta K/K_m\) and expand the first-order terms in these parameters in powers of \(\psi\) around \(\psi = 0\). We solve for the correction \(\beta^1(\psi)\) with boundary conditions \(\beta^1(0) = 0\) and \(\beta^{1\prime}(0) = 0\), which respectively enforce that our approximate solution is exact at the equator and that the correction obeys the up-down symmetry of the spindle. To leading order in \(\psi\), our perturbation solution is therefore
	\begin{equation}
		\beta^{\epsilon}(\psi) = \beta_0 -\frac{\Delta K}{K_m} \frac{\psi^2 \sin{2\beta_0}}{2(1+\cos{\eta_0})}.
	\end{equation}	
From this, we observe that our leading order correction is quadratic in \(\psi\). If \(\Delta K\) is positive the correction causes the twist angle to decrease from the value of \(\beta_0\) away from the equator while a negative \(\Delta K\) causes the twist angle to increase. Experimentally, the value of \(\Delta K/K_m\) depends on the material. As an example, for 5CB \(\Delta K/K_m\) takes a value in the range 0.1-0.14~\cite{Bogi2001, Zakharov2002}, which, as we will demonstrate, can be considered sufficiently small that our perturbation solution is a reasonable approximation to the exact solution.

Near to the spindle tip, the divergence of the \(\sec{\psi}\) term forces the \(\sin{2\beta(\psi)}\) term towards zero. Expanding \cref{eq:eeq-beta(psi)} near to the spindle tips, we find that the dominant term is 
	\begin{equation}
		-\frac{\Delta K \sin{\eta_0}\sin{2\beta(\psi)}}{\pm\frac{\pi}{2}\mp\psi},
	\end{equation}	 
where the plus and minus respectively correspond to the tips at \(\pm \pi/2\). When \(|\pm\pi/2\mp\psi| \lesssim |\Delta K \sin{\eta_0}| \) this term drives \(\sin{2\beta(\psi)}\) towards zero. Based on our solution near to the equator, we would therefore expect that if \(\Delta K > 0\) this drives the twist angle towards zero while if \(\Delta K<0\) the twist angle tends towards \(\pi/2\). However, the \(\gamma\) dependent term in \cref{eq:eeq-beta(psi)} diverges as the twist angle tends towards \(\pi/2\). We therefore expect that for \(\Delta K < 0 \) the twist angle will initially increase towards \(\pi/2\) before dropping off to zero very close to the tips.

\begin{figure}
		\centering
		\includegraphics[width = 0.49\textwidth]{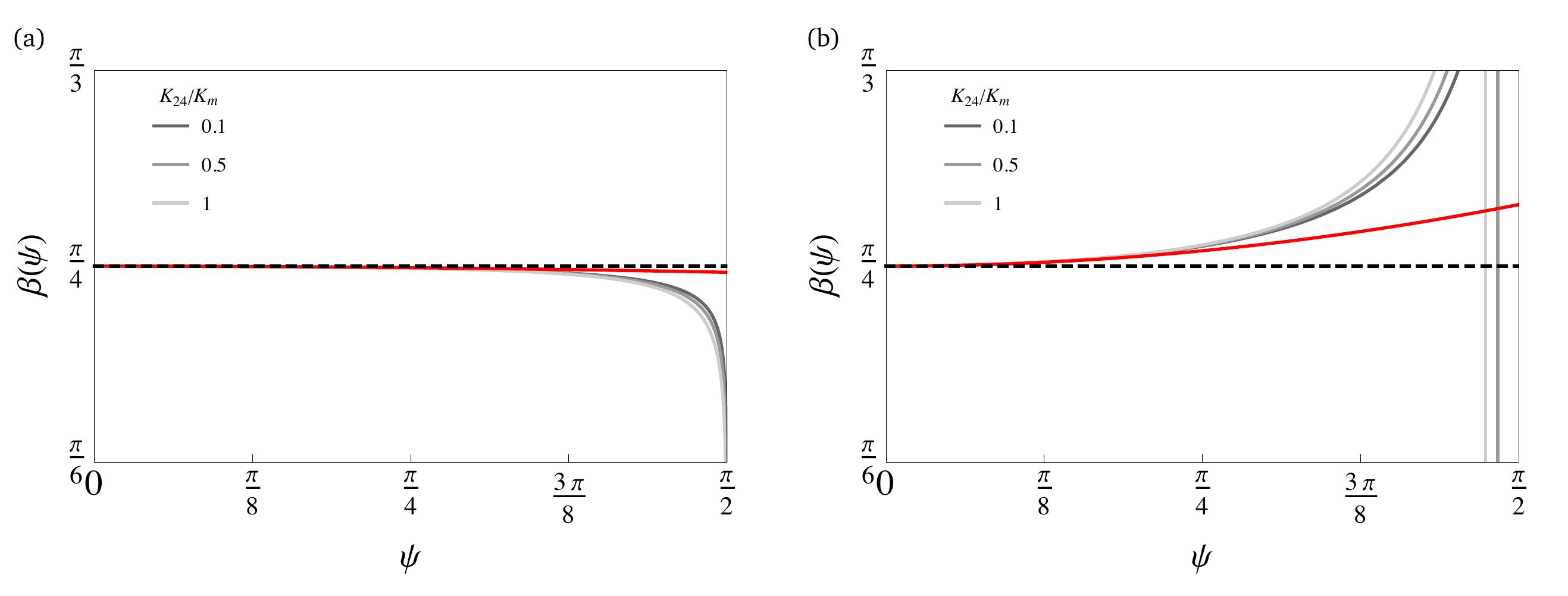}
		\caption{Solutions for the twist angle \(\beta(\psi)\), with \(\beta_0 = \pi/4\), \(\eta_0 = \pi/3\) and \(\gamma\) defined by \cref{eq:cos3beta} for (a) \(\Delta K/K_m = 0.01\) and (b) \(\Delta K/K_m = -0.1\). Solid gray lines correspond to numerical solutions for the values of \(K_{24}\) shown in the plot legend. The dashed line represents an ideal loxodrome solution while the red curve corresponds to our asymptotic solution \(\beta^{\epsilon}(\psi)\).}	
		\label{fig:first-correction-psi}
		\end{figure}

To assess the validity of this solution, we numerically solve \cref{eq:eeq-beta(psi)} with initial conditions \(\beta(0) = \beta_0\) and \(\beta'(0) = 0\), which we expect due to up-down symmetry of the spindle. We choose the twist angle at the equator \(\beta_0 = \pi/4\), which maximizes the contribution of the \(\sin{2\beta_0}\) term in the correction, \(\eta_0 = \pi/3\), which corresponds to a spindle with aspect ratio \(\sqrt{3}\), and \(\gamma\) from the relation in \cref{eq:cos3beta}. We plot the numerical solutions in \cref{fig:first-correction-psi} for \((a)\) \(\Delta K/K_m = 0.01\) and (b) \(\Delta K/K_m = -0.1\). Numerical solutions are plotted for a range of \(K_{24}\) values along with the ideal loxodrome \(\beta_0\) and our perturbation solution at the equator \(\beta^{\epsilon}(\psi)\). Near to the equator, \(\beta^{\epsilon}(\psi)\) gives a very good approximation to the numerical solution, with the smaller magnitude \(\Delta K\) value giving a good approximation for a larger range of \(\psi\) values. As expected, the solution with \(\Delta K > 0 \) causes the numerical solution for the twist angle to decrease away from \(\beta_0\) while \(\Delta K < 0 \) causes the opposite behavior before all of the numerical solutions diverge at the tips. We observe that the numerical solutions are only weakly dependent on the value of \(K_{24}\), which is consistent with the leading behavior depending on its value only though \(\beta_0\). When \(\Delta K/K_m = -0.1\), the numerical solutions are all within \(5^{\circ}\) of the value of \(\beta_0\) up to \(\tilde\psi = \psi/(\pi/2) =  0.7\), meaning that the twist angle is within this tolerance of the loxodrome twist angle over \(85\%\) of the surface area. Decreasing the magnitude of the perturbation \(\Delta K/K_m\) to \(0.01\) results in the twist angle being within \(5^{\circ}\) of \(\beta_0\) up to \(\tilde\psi = 0.97\) and within \(1^{\circ}\) of \(\beta_0\) up to \(\tilde\psi = 0.82\). We therefore conclude that away from the spindle tips, the loxodrome solution is a valid approximation for the twist angle if \(\Delta K/K_m\) is small.

\section{Loxodromes on a general surface of revolution}
Having established the conditions under which twisted loxodromes are exact or approximate solutions on the spindle surface, we now turn our attention to establishing conditions on the shape of a surface for it to support such a loxodrome solution. We consider a general surface of revolution, described in cylindrical coordinates \((\rho,\phi, z)\). The surface is formed by revolving its shape profile \(\rho(z)>0\) about the \(z\) axis. We once again introduce a general director field 
	\begin{equation}
		\vb{n} = \cos{\beta(z)}\vu{e}_v+\sin{\beta(z)}\vu{e}_{\phi}
	\end{equation}
that lies tangent to the surface and follows the azimuthal symmetry of the surface. As before, \(\vu{e}_{\phi}\) is a unit vector in the azimuthal direction while \(\vu{e}_{v}\) is a unit vector tangential to lines of constant \(\phi\), equivalent to \(\vu{e}_{\psi}\) in the bispherical coordinates. We again calculate the free energy \(F = F_F + F_l\) for our surface and minimize for \(\beta(z)\). A loxodrome solution, \(\beta(z) = \beta_0\) and \(\beta'(z) = \beta''(z) = 0\) leads us to require that 
			\begin{align}
			0 &= -K_{24}\rho(z)[\kappa_{\nu}(z)-\kappa_{\phi}(z)]\sin{2\beta_0}+ \frac{\gamma}{2\pi}\tan{\beta_0}\sec{\beta_0} \nonumber\\
				& - \Delta K \left[\frac{1}{\rho(z)}+\rho(z)(\kappa_{\nu}(z)-\kappa_{\phi}(z))\right]\sin{2\beta_0}.
				\label{eq:eeqz}
		\end{align}
	Here we have introduced \(\kappa_{v}(z)\) and \(\kappa_{\phi}(z)\), the principal curvatures of the surface, which depend on \(\rho(z)\) and its derivatives. On the spindle surface, we found an exact twisted loxodrome solution when \(\Delta K = 0\). In this case the coefficient of the \(K_{24}\) term was constant and the loxodrome solution resulted in the balance of this term with the \(\gamma\) term. Imposing the same conditions here, we therefore need to determine the conditions under which \(\rho(z)[\kappa_{\nu}(z)-\kappa_{\phi}(z)]\) takes on a constant value \(\xi\).

	 Solving for \(\xi\) with the requirement that \(\rho(z) = \rho(-z)\), we first find that cylinders, for which \(\rho(z)\) is constant, and conical surfaces, for which \(\rho(z)\) is linear in \(z\) satisfy this condition. The most general shape profile for which \(\xi\) is constant satisfies \((\rho \pm R_1)^2 + z^2 = R_2^2\). This is the equation of a circle of radius \(R_2\) centered at \((\pm R_1,0)\) in the \((\rho,z)\) plane. The constants \(R_1\) and \(R_2\) can be taken to be non-negative without loss of generality. The constant \(\xi = R_1/R_2>0\) defines the shape profile of the surface formed by revolving the section with \(\rho(z)>0\). We note that a spherical surface, for which \(\xi = 0\), does not support our twisted loxodrome solution because it causes the \(K_{24}\) term to vanish.
	 
\begin{figure}
	 \centering
	 \includegraphics[width=0.49\textwidth]{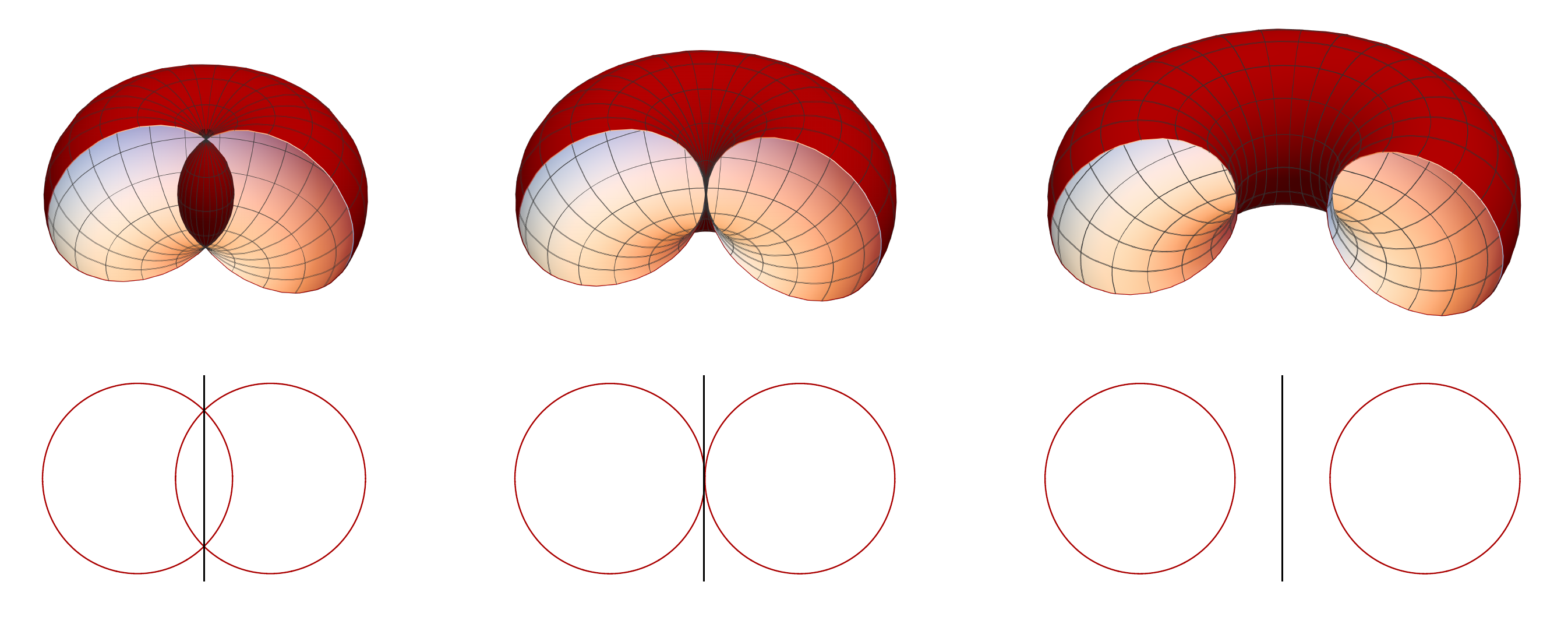}
	 \caption{Schematics of surfaces and cross sections of the general torus. From left to right: spindle torus, horn torus and hole torus. The black vertical lines indicate the axis of revolution.}
	 \label{fig:torus}
	 \end{figure}

	 The general shape profile we have derived generates surfaces of the standard torus, as depicted in \cref{fig:torus}. If \(\xi<1\) the circles forming the shape profile cross the \(z\)-axis and generate a spindle torus.  Taking the \(\rho(z)>0\) arcs of the generating circles, the plus sign solution generates an apple surface, which corresponds to the outer surface of the spindle torus, while the minus sign generates the spindle surface we have previously considered.  In both of these cases the twist angle obeys an equivalent expression to \cref{eq:cos3beta}. When \(\xi>1\), only one of the generating circles lies within the \(\rho(z)>0\) region and the resulting surface is a hole torus while the case \(\xi=1\) gives the limiting case of a horn torus. 
	 
	On the spindle and apple surfaces, it is natural for the fixed length of the loxodrome to span the range of \(z\) values between the two singularities on the surface. This means that for a fixed length curve, the twist angle of the loxodrome is uniquely determined by the value of \(\xi\) on the surface. By contrast, for the hole torus, cylinder and cone there is not a naturally defined range of \(z\) values that the loxodrome should span. For given properties of these surfaces, a choice of the span of \(z\) must be chosen to determine the loxodrome twist angle. On the hole torus, special choices of twist angle allow a closed loop to form, which therefore allows a clear length constraint to be defined. 
		 
We now restrict ourselves to spindle-like surfaces and consider the effect of shape perturbations on the stability of the loxodrome solution. To do this, we construct a generalized spindle for which the shape profile is now an arc of an ellipse instead of a circle. As is the case for the spindle generated from the arc of a circle, which we refer to as an ideal spindle, the generalized spindle is symmetric upon reflection in the \(z\)-axis and has pointed tips except in the case of a perfect sphere. The generalized spindle has aspect ratio \(u_0\), defined so that \(u_0 = 1\) corresponds to a perfect sphere and our prolate spindles have \(u_0>1\). The profile curve has eccentricity \(0\leq e < 1\), where \(e = 0\) corresponds to a perfect circle and \(e = 1\) corresponds to a parabola. A general expression for the shape profile \(\rho(z)\) of the generalized spindle is therefore						
		\begin{equation}
		\small
				\rho(z)=-\frac{u_0^2-1+e^2}{2(1-e^2)} + \frac{\sqrt{(1-e^2+u_0^2)^2-4(1-e^2)z^2}}{2(1-e^2)}.
		\end{equation}		
		We examine perturbations away from the ideal spindle by treating \(e\) as a small parameter and expanding \cref{eq:eeqz} in powers of \(e\) with \(\Delta K = 0\). We introduce a perturbation expansion for the twist angle \(\beta^{\epsilon}(z) = \beta^0(z) + \epsilon \beta^1(z)\), where \(\epsilon\) is a again a small parameter that we identify with \(e^2\), the lowest nonzero power of \(e\) in our expansion. 
		
		We find that the leading order expression allows for loxodromes of constant angle \(\beta_0\) with the condition	
		\begin{equation}
				\cos^3{\beta_0}  = \frac{\gamma(u_0^2+1)}{4\pi K_{24}(u_0^2-1)}.
		\end{equation}	
		In the ideal spindle, \(\cos{\eta_0} \equiv (u_0^2-1)/(u_0^2+1)\), so this expression is equivalent to \cref{eq:cos3beta}. We calculate the correction term by Taylor expanding the contributing terms to leading order in \(z\) and using boundary conditions that again set \(\beta^1(0) = 0\) and \(\beta^{1'}(0) = 0\). Our perturbation solution therefore becomes
		\begin{equation}
			\beta^{\epsilon}(z) = \beta_0 + e^2 \frac{K_{24}\sin{2\beta_0}}{K_m(1+u_0^2)^2}z^2.
		\end{equation}	
		Given that the constants multiplying \(e^2\) result in an expression that is order one, we find that the correction is quadratically small in the correction to the constant loxodrome solution. We therefore find that the loxodrome solution is stable to small perturbations away from the ideal circle-arc spindle with \(e = 0\).

\section{Loxodrome solutions in the bulk}

We now turn our attention to bulk twisted structures enclosed within a spindle-shaped surface parameterized by \(\eta = \eta_0\). Following prior investigations of bipolar spindles, we take the director field inside the spindle to adopt a bispherical configuration in which the director field is tangential to surfaces of constant \(\eta\) in the bispherical coordinate system~\cite{Williams1985-028, Kaznacheev2002, Kaznacheev2003, Prinsen2003, Prinsen2004, Prinsen2004b}. We consider only truly bispherical structures in which the boojum defects of the nematic sit at the spindle tips, which we justify based on the observed structures of our polymer particles~\cite{Ansell2019}. Also in line with prior investigations~\cite{Williams1986, Prinsen2004}, we assume that in twisted bipolar configurations, the twisting maintains the bispherical structure and therefore occurs within the local tangent plane to the surface of constant \(\eta\) upon which any point in the bulk resides.

Starting from these assumptions, Williams~\cite{Williams1986} investigated the twisting behavior of a nematic confined to a spherical bipolar droplet with strong planar anchoring by assuming that the twist angle of the nematic followed loxodromes on surfaces of constant \(\eta\) and could be parameterized in terms of a function \(\beta_0(\eta)\). Using the ansatz \(\beta_0(\eta)\propto \sin{\eta}\), which satisfies the key requirements of the solution, Williams showed that a twisted bipolar configuration lowers the Frank free energy of the system if the elastic constants obey the inequality \(K_3 \lesssim 2.32 (K_1 -K_2)\). Prinsen and van der Schoot~\cite{Prinsen2004} then used these same assumptions to extended this inequality to bipolar and quasi-bipolar spindle-shaped systems.

This twisted loxodrome structure does not analytically minimize the Frank free energy of the system. In order to verify the validity of the loxodrome assumption, we therefore investigate numerical solutions for the twist angle \(\beta(\eta,\psi)\) of a director field constrained to be tangential to the bispherical structure. We use the MATLAB PDE Toolbox\texttrademark~\cite{Matlab} to solve for the twist angle in the \(\psi>0\) region of the spindle and determine the \(\psi<0\) solution using the inversion symmetry of the spindle. We set the boundary conditions \(\beta(0,\psi) = 0\), which is required to ensure that no defects are present along the central axis of revolution, and the derivative \(\beta_{\psi}(\eta,0) = 0\) to ensure the solution obeys the inversion symmetry of the spindle. We have more freedom with the choice of the boundary conditions on the spindle surface. Given that we are interested in the twisted loxodrome structure, we impose that on the surface the twist pattern follows a loxodrome with chosen twist angle \(\beta_0 \) such that \(\beta(\eta_0, \psi) = \beta_0\).

\begin{figure}
	\centering
	\includegraphics[width=0.49\textwidth]{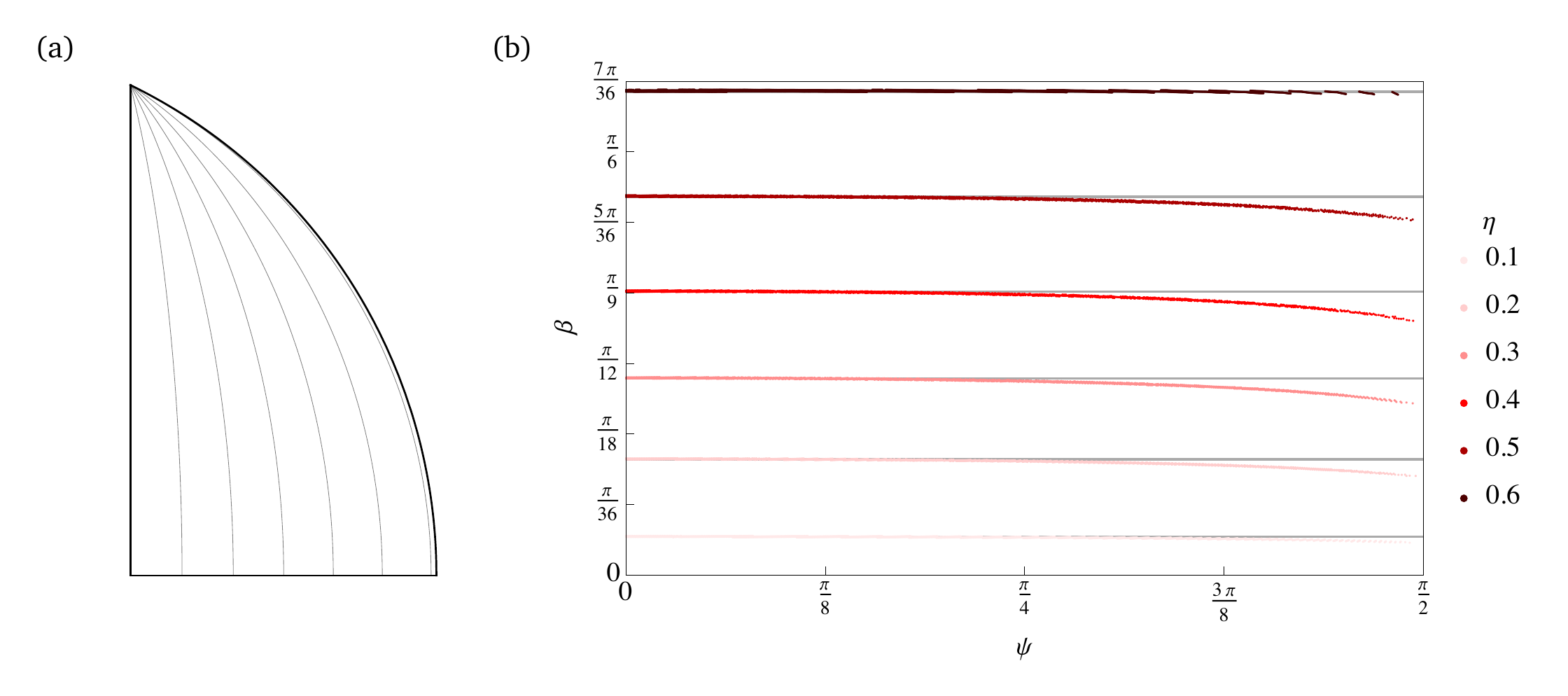}
	\caption{(a) Contours of constant \(\beta\) in the numerical solution for the bulk twist angle  \(\beta(\eta,\psi)\) for a spindle with size \(\eta_0 = 1.18\) and surface twist angle \(\beta(\eta_0,\psi) = 7\pi/36\). (b) Slices through this solution along surfaces of constant \(\eta\). The colored lines show the numerical solutions while the gray 	lines indicate the solution \(\beta(\eta,0)\), corresponding to the expected loxodrome twist angle at a given \(\eta\) value.}
	\label{fig:bulk-numsol}
	\end{figure}

\Cref{fig:bulk-numsol}(a) shows contours of constant twist angle in the numerical solution for \(\beta_0 = 7\pi/36\)  on a spindle with aspect ratio \(u_0 = 1.6\) (\(\eta_0 = 1.18\)) in the one-constant approximation, which appear consistent with a twisted bipolar structure. In \cref{fig:bulk-numsol}(b) we therefore show how the twist angle of this solution varies with \(\psi\) on surfaces of constant \(\eta\). In this plot, a loxodrome twist pattern corresponds to a horizontal line. We observe that the solutions are very close to being straight lines over much of the range of \(\psi\). Near to the spindle tips, we observe that the twist angle decreases on each of the surfaces, with the most deviation for the values of \(\eta\) furthest from those at which the boundary conditions are imposed. The decrease in twist angle near to the tip is likely due to the system trying to mitigate the large twist energy in this region. Our results verify that the twisted loxodrome solution is a good approximation to the internal twist structure. 

We analyze the bulk Frank free energy of these numerical solutions for different values of \(u_0\), \(\beta_0\) and the ratio of elastic constants \(K_2/K\), where \(K = K_1 = K_3\). As expected, our numerical results are consistent with the scaling analysis of Prinsen and van der Schoot~\cite{Prinsen2004}. That is, at aspect ratios close to one at the smaller \(K_2/K\) values there is a minimum in the free energy at a non-zero twist angle while at higher \(K_2/K\) values the untwisted configuration minimizes the free energy. At larger \(u_0\) values the bound on \(K_2/K\) at which twisting is preferred decreases until it reaches zero at some critical value above which the system remains untwisted. 

While our numerical results are consistent with the scaling theory, they do not capture the observed behavior of our twisted polymer system. In the experimental system, there is a critical aspect ratio for the onset of twisting above which the twist angle increases with aspect ratio thereby displaying the opposite trend to the twisting behavior expected in a system governed by the Frank free energy alone. As was the case when considering a spindle-shaped surface, we therefore seek to incorporate additional terms into the free energy to account for the polymer nature of the system under consideration. In line with our results on the spindle surface and our previous geometric model~\cite{Ansell2019}, we once again impose a length constraint term in addition to the Frank free energy.

We take a geometric approach to incorporating the length constraint condition and then examine the optimal twist pattern of the resulting free energy. We now assume that that the twist pattern follows loxodromes on surfaces of constant \(\eta\), which we justify using the numerical results presented in \cref{fig:bulk-numsol}. In our previous geometric model of the surface twisting~\cite{Ansell2019}, the length of the meridians of the spindle surface at the critical aspect ratio gave us fixed length of the twisted loxodrome curves at larger aspect ratios (smaller volumes). We therefore model the bulk structure by mapping the length of meridians on internal surfaces of constant \(\eta\) enclosed within a spindle at the critical aspect ratio to the length of a twisted loxodrome on a surface of constant \(\eta\) in a more elongated twisted spindle.

\begin{figure}
	\centering
	\includegraphics[width=0.45\textwidth]{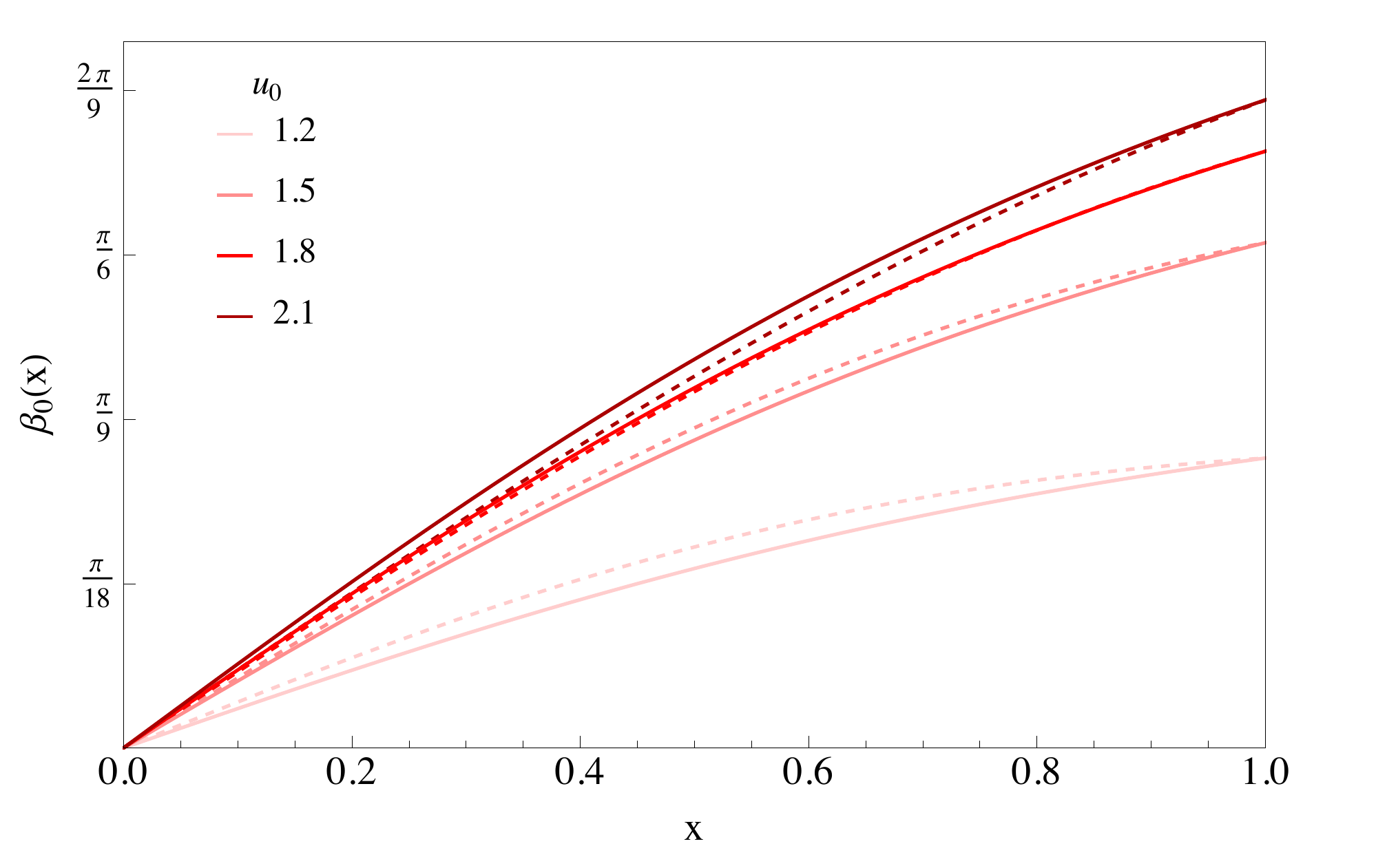}
	\caption{Plots of the expression in \cref{eq:betax} showing the predicted loxodrome twist angle in the bulk spindle for \(u^* = 1.1\) and a range of \(u_0\) values (solid lines). The dashed lines correspond to the expression \(\beta(\eta)\propto\sin(\eta)\) with the proportionality constant chosen to match the twist angle at the surface in each case.}
	\label{fig:internal-twist}
	\end{figure}

In order to construct the mapping, we introduce the fractional distance \(x\) along the minor axis of a spindle between its central axis and outer surface. The surfaces of constant \(\eta\) that form the bulk structure can therefore all be ascribed an \(x\) value. The length of a meridian on a surface at a given \(x\) value in the spindle with critical aspect ratio \(u^*\) becomes the length of the twisted loxodrome at the same \(x\) values in the smaller twisted spindle with aspect ratio \(u_0\). If the length of a meridian is \(l_m\), the length of a loxodrome with twist angle \(\beta_0\) is \(l_m \sec{\beta_0}\). Our mapping can therefore be expressed as \(l_m(u^*, x) = l_m(u_0,x)\sec{\beta_0(x)}\), which leads to the loxodrome twist angle satisfying
	\begin{equation}
		\cos{\beta_0(x)} = \frac{u^*(x^2+u_0^2)\tan^{-1}\qty(x/u_0)}{u_0(x^2+u^{*2})\tan^{-1}(x/u^*)}.
		\label{eq:betax}
	\end{equation}
	Plots of this expression for a spindle with \(u^*=1.1\) are shown in \cref{fig:internal-twist} (solid lines) for a range of \(u_0\) values. The expression gives the general behavior we would expect from the twist angle in that there is no twisting at the center (\(x=0\)) and the twist angle monotonically increases to a maximum value on the spindle surface. The relation is also consistent with the approximate solution \(\beta(\eta)\propto\sin{\eta}\) used in prior investigations~\cite{Williams1986, Prinsen2004}, as shown in the dashed lines in \cref{fig:internal-twist} for which the proportionality constant has been chosen to match the twist angle at the surface. We note that in this mapping we have fixed the major axis length of the spindle, meaning that changes in aspect ratio are as a direct result of changes in minor axis length. Setting \(x=1\) therefore gives the idealized behavior of the geometric model we developed for our experimental polymer system~\cite{Ansell2019}, in which we had to adapt the model to account for an amount of length reduction in the major axis due to polymer chain folding.
	
	From our bulk geometric model, we construct a new total bulk free energy \(F = F_b + F_l\), where \(F_b\) is the bulk part of the Frank free energy. We consider the influence of the saddle-splay energy on the behavior of the model later in this section. The constraint term \(F_l\) takes the form
	\begin{equation}
		F_l = \int \dd \eta \gamma(\eta)\qty[l_m(\eta_0,\eta)\sec{\beta_0(\eta)}-l_m(\eta^*,\eta)],
	\end{equation}
	where \(\gamma(\eta)\) is a Lagrange multiplier that constrains the lengths on each surface of constant \(\eta\) and we convert \(x\) values into \(\eta\) values in the final spindle using \(x = \tan(\eta/2)\cot(\eta_0/2)\). The constraint term ensures that the twist profile is given by the expression in \cref{eq:betax} while \(\gamma(\eta)\) can be determined by solving the Euler-Lagrange equations of the free energy.

\begin{figure}
	\centering
	\includegraphics[width=0.45\textwidth]{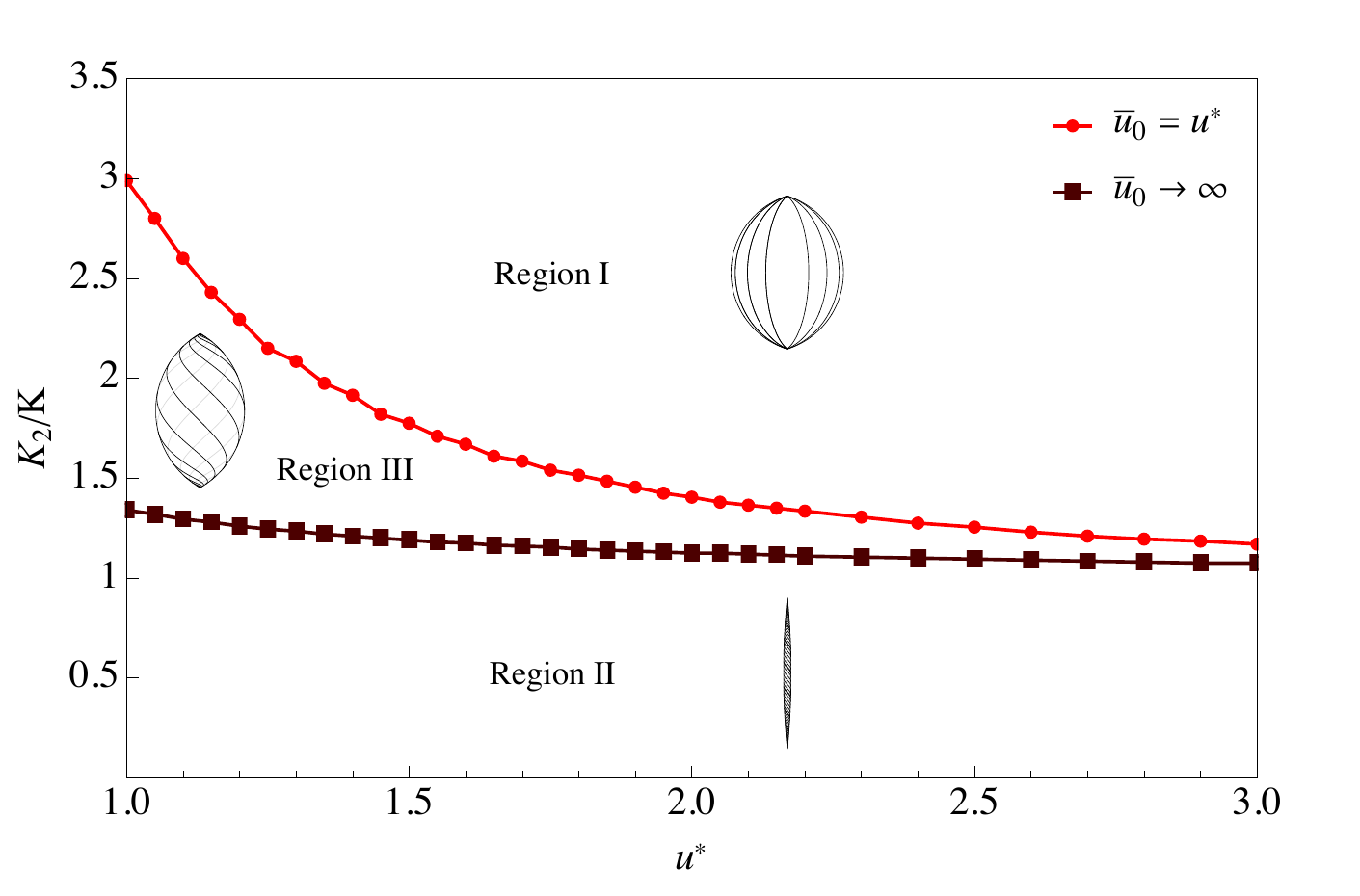}
	\caption{Optimal spindle configurations in the \(K_2/K\)-\(u^*\) parameter space. In region I, the untwisted configuration (\(\bar{u}_0\to u^*\)) is optimal while in region II the spindle twists as much as possible (\(\bar{u}_0\to\infty\)). Region III is the intermediate region in which the optimal aspect ratio takes on a value between these two limiting cases. The plotted points indicate the numerically determined boundaries of each region.}
	\label{fig:bulk-twist}
	\end{figure}

	We explore the behavior of the free energy as a function of the twisted spindle aspect ratio \(u_0>u^*\) for different values of \(u^*\) and \(K_2/K\), with a cut off \(\eta_{\text{min}} = 10^{-6}\) at the center of the spindle to prevent the free energy from diverging. We observe that there is always a single minimum in the free energy at an optimal aspect ratio \(\bar{u}_0\), the value of which allows us to classify the expected behavior of such a system into one of three regimes. The first case is that the minimum in the free energy occurs when \(\bar{u}_0 = u^*\), meaning that the optimal configuration is for the spindle to remain untwisted. This regime is observed above a particular value of \(K_2/K\) that depends on the value of \(u^*\). In the second case, which occurs when \(K_2/K\) is below a particular value, the optimal aspect ratio is \(\bar{u}_0\to\infty\) meaning that the free energy wants the system to twist as much as possible. In a real system, the physical bulk of the material would prevent the system from twisting this far and would have an effect in determining the optimal aspect ratio. In the final case there is a minimum in the free energy at a finite value of \(\bar{u}_0\), resulting an an optimal aspect ratio and therefore optimal twist angle in the system.  \Cref{fig:bulk-twist} shows the regions of the \(K_2/K\)--\(u^*\) parameter space for which the optimal twist behavior falls into each of these three regimes.

	Comparing the twisting behavior of our new free energy that incorporates the constraint condition to that of the bulk Frank free energy alone, we observe that including the additional term results in twisting being favorable over a wider parameter range. In a spherical bipolar system, Williams~\cite{Williams1986} showed that if \(K_1 = K_3 = K\) the Frank free energy allows a twisted loxodrome solution if \(K_2/K < 0.57\). Following on from this, Prinsen and van der Schoot~\cite{Prinsen2004} showed that this bound is highest for spherical tactoids and that increasing the aspect ratio decreases the maximum value of \(K_2/K\) at which twisting is preferable. By contrast, in our formulation we observe that, while the maximum value of \(K_2/K\) at which twisting can occur varies with the reference aspect ratio \(u^*\), when \(u^* = 1.0\) twisting can occur up to \(K_2/K \lesssim 3.0\) and this value gradually reduces to \(K_2/K\lesssim 1.17\) when \(u^*=3\). In particular, including the length constraint condition means that in the one-constant approximation \(K_2/K = 1\), we would expect to observe a twisted configuration in our system.
	
Having considered the influence of the bulk elasticity on the twisting behavior of our system, we now turn our attention to the influence of the saddle-splay energy on the optimal spindle structure. The saddle-splay energy is a surface term, and therefore depends only on the value of the twist angle at the surface. The expression for the saddle-splay is given by \(F_{SS} = -4 \pi  K_{24} \eta_0 \cot{\eta_0} \sin^2{\beta_0(1)} \), where \(\beta_0(x)\) is defined in \cref{eq:betax} and we have chosen the zero-point of the energy such that the saddle-splay contribution is zero when there is no twisting. We know that the saddle-splay elastic constant obeys \(\abs{K_{24}} \leq 2 K\). If we assume the value of \(K_{24}\) is of the same order of magnitude as the bulk elastic constants, the contribution from the saddle-splay energy is of the same order of magnitude as the bulk free energy. 

\begin{figure}
\centering
\includegraphics[width = 0.45\textwidth]{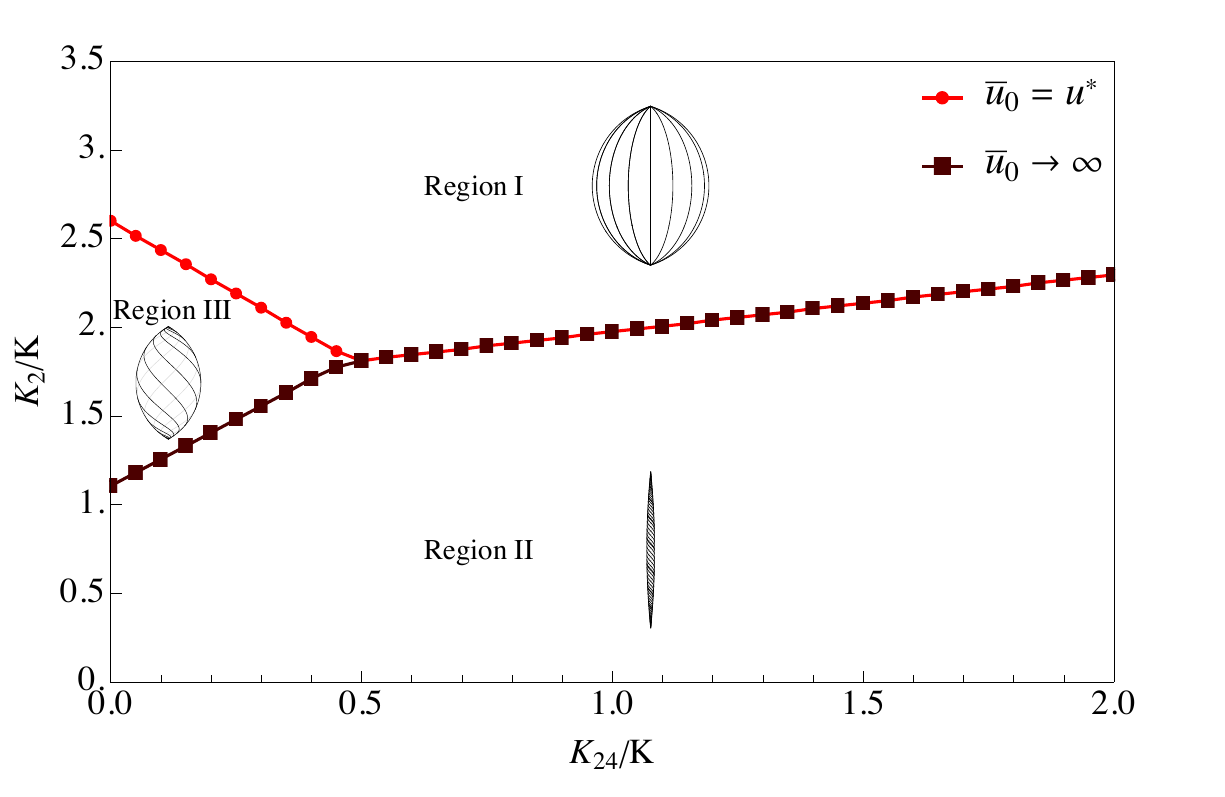}
\caption{Optimal spindle configurations for different values of the saddle-splay constant \(K_{24}/K\)  for a spindle with \(u^* = 1.1\). As in \cref{fig:bulk-twist}, region I corresponds to the optimal structure being the untwisted configuration \(\bar{u}_0  = u^*\), region II corresponds to the optimal structure being \(\bar{u}_0\to\infty \), and in region III, which exists only at smaller \(K_{24}/K\) values, the optimal structure is between these two limits.}
\label{fig:ss-twist}
\end{figure}

Given that we expect \(K_{24}\) to be positive, the saddle-splay energy favors a highly twisted configuration. We observe that including the saddle-splay term in our model raises the bound on \(K_2/K\) at which the system becomes as twisted as possible  (\(\bar{u}_0\to\infty\)). The bound on \(K_2/K\) at which the system is able to adopt a twisted configuration initially decreases as \(K_{24}/K\) increases. Above some critical value of \(K_{24}/K\), the intermediate regime \(u^*<\bar{u}_0<\infty\) no longer exists and a single curve separates the untwisted and maximally twisted regions in the \(K_{24}/K\)--\(K_2/K\) parameter space. \Cref{fig:ss-twist} shows the influence of \(K_{24}/K\) on the twisting behavior for a spindle with reference aspect ratio \(u^* = 1.1\) in which we observe that the intermediate regime does not exist for \(K_{24}/K>0.5\). Across reference aspect ratios explored in the range \(1\leq u^* \leq 3\) we observe the same general behavior as observed in \cref{fig:ss-twist} with the magnitude of the critical aspect value of \(K_{24}/K\) increasing with \(u^*\) up to \(K_{24}/K = 0.9\) at \(u^* = 3\). We therefore find that the saddle-splay term does indeed make the twisted structure more preferable within our system.

\section{Conclusions}
We have investigated the twisting behavior of spindle-shaped systems with bipolar nematic ordering, with a particular focus on developing a model of liquid crystalline polymer particles that display twisting behavior during deswelling. We have demonstrated that including a constraint on the length of the integral curves of the system produces a model that captures the behavior of this system, thereby providing an energetic pathway to the twisted loxodrome pattern that we previously derived using a geometric approach~\cite{Ansell2019}. It is the inclusion of this length constraint that results in the model predicting remarkably different behavior to that expected for a typical nematic system that is governed by the Frank free energy alone.

We demonstrated that an exact twisted loxodrome solution minimizes our total free energy on a spindle-shaped surface, subject in the one-constant approximation (\(K_1 = K_3)\). Allowing small deviations away from this condition, or small perturbations in the shape-profile of the spindle, leads to solutions for which the twisted loxodrome is the leading order term and the correction terms are small. In a bulk bipolar nematic confined to a spindle-shaped region, we showed that if the twist pattern on the surface follows loxodromes then the bulk twisting structure is well approximated by twisted loxodromes on surfaces of constant \(\eta\). By developing a geometric model of this twisted structure in which the loxodrome twist angle on each surface of constant \(\eta\) is determined by a length constraint condition, we have formulated a model for which twisting behavior is optimal in the system over a larger parameter range than in a nematic system governed by the Frank free energy alone. Crucially, the model we developed in section 5 also captures the shape change and twisting behavior of the polymer system~\cite{Ansell2019} and we look forward to future exploration of the internal structures of these spindles to test the predicted model.

\section*{Conflicts of interest}
There are no conflicts to declare.

\section*{Acknowledgements}
H.S.A. and R.D.K. were supported by NSF MRSEC Grant DMR-1720530 and a Simons Investigator Grant from the Simons Foundation to R.D.K.







%

\end{document}